\documentclass[12pt]{article}

\usepackage[margin=1.5cm]{geometry}

\usepackage{mathtools}
\usepackage{microtype}
\usepackage{hyperref}

\usepackage{etex}
\usepackage[subpreambles=true]{standalone}
\usepackage[utf8]{inputenc}
\usepackage[english]{babel}
\usepackage{import}
\usepackage{graphicx}

\usepackage{amsmath,amssymb,mathrsfs,graphicx,indentfirst,amsfonts,listings}
\usepackage{subfig}
\usepackage{float}
\usepackage{verbatim}
\usepackage{bbm}
\usepackage{hyperref}
\usepackage{caption}
\usepackage{color}
\usepackage{tikz}
\usepackage{tikz-qtree}

\usepackage{tabularx}

\begin{document}


\title{Modelling gene content across a phylogeny to determine when genes become associated~\thanks{We would like to thank the Australian Research Council for funding this research through Discovery Project DP180100352.}}
 

	\author{
		Jiahao Diao\thanks{Discipline of Mathematics, University of Tasmania, Australia, email: jiahao.diao@utas.edu.au}
		\ and
		Ma{\l}gorzata M. O'Reilly~\thanks{Discipline of Mathematics, University of Tasmania, Australia, email: malgorzata.oreilly@utas.edu.au}
		\ and
		Barbara R. Holland\thanks{Australian Research Council Centre of Excellence for Plant Success, Discipline of Mathematics, University of Tasmania, Australia, email: barbara.holland@utas.edu.au} 
	}

\date{\today}
\maketitle
	
\section{Introduction}	
In this work, we develop a stochastic model of gene gain and loss with the aim of inferring when (if at all), in evolutionary history, an association between two genes arises. The data we consider is a species tree along with information on the presence or absence of two genes in each of the species. The biological motivation for our model is that if two genes are involved in the same biochemical pathway, i.e. they are both required for some function, then the rate of gain or loss of one gene in the pathway should depend upon the presence or absence of the other gene in the pathway. However, if the two genes are not functionally linked, then the rate of gain or loss of one gene should be independent of the state of the another gene.

Our model builds upon some existing models. Barker et al. \cite{barker2007constrained} were the first to note that the presence and absence of two genes can be modelled as a four-state Markov process. The states $\{00, 01, 10, 11\}$ record each gene's presence  (1) or absence (0). Each transition can only have one gene loss or gain (e.g. $00\to01$ or $00\to10$ are permitted but not $00\to11$). To test if two genes were associated they used likelihood ratio tests to compare a model where substitution rates were consistent with independent evolution (e.g. $q_{00\to01}  = q_{10\to11}  $) versus a model where rates were dependent. Marazzi et al.~\cite{marazzi2012locating} proposed a precursor model for the evolution of a binary trait. Instead of transiting directly from $0\to1$ or $1\to0$, they developed a model with an intermediate unobservable state. Boyko et al. \cite{boyko2021generalized} further generalised this idea to models with hidden rate classes. For example, two different rate classes may have the same observable states but different transition rates among the states.

Our model combines the dependent and independent models considered by Barker et al. \cite{barker2007constrained}  as two hidden rate classes in the framework proposed by Boyko et al. \cite{boyko2021generalized}.

We simulate data under this model using the R package corHMM from \cite{boyko2021generalized}. We aim to determine under what conditions a shift from the independent rates class to the dependent rates class can be detected. For example, how large a tree is required and how large a shift in the rates is needed before Akaike information criterion (AIC) supports a model with two rate classes over a simpler model with just one rate class? If a model with two rate classes is preferred, can it correctly detect where on the evolutionary tree the shift occurred?


\section{Neofunctionalisation model I}\label{sec:model1}

First, we propose the following continuous-time Markov chain $\{X(t):t\geq 0\}$ with state space
\begin{eqnarray}
\mathcal{S}&=&\{00,01,10,11,\widetilde{00},
\widetilde{01},\widetilde{10},\widetilde{11}\},
\end{eqnarray}
where $1$ and $0$ corresponds to the presence or absence of a gene, states in the subset $\mathcal{S}_{ind}=\{00,01,10,11\}\subset \mathcal{S}$ correspond to genes that do not depend on one another, and states in the subset $\mathcal{S}_{d}=\{\widetilde{00},
\widetilde{01},\widetilde{10},\widetilde{11}\}\subset \mathcal{S}$ correspond to genes that depend on one another. As example, state $X(t)=\widetilde{11}$ means that both genes are present in the species, and are also dependent on one another, at time $t$.

We assume that the transition rates of this process are collected in the generator matrix ${\bf A}=[{A}_{i,j}]_{i,j\in\mathcal{S}}$ given by,
\begin{align}\label{transi1}
{\bf A} =
\begin{array}{cc}&
\begin{array}{cccccccc}
00 & 01 & 10 &11 &\quad \widetilde{00} &\widetilde{01} &\widetilde{10} &\widetilde{11}\\
\end{array}\\
\begin{matrix}
{00} \\{01} \\{10} \\{11}\\
\widetilde{00} \\\widetilde{01} \\\widetilde{10} \\ \widetilde{11}
\end{matrix}&
\begin{bmatrix}[cccc|cccc]
*&q_{12}&q_{13}&0&q_1&0&0&0\\
q_{21}&*&0&q_{24}&0&q_2&0&0\\
q_{31}&0&*&q_{34}&0&0&q_3&0\\
0&q_{42}&q_{43}&*&0&0&0&q_4\\
\hline
q&0&0&0&*&\widetilde{q_{12}}&\widetilde{q_{13}}&0\\
0&q&0&0&\widetilde{q_{21}}&*&0&\widetilde{q_{24}}\\	0&0&q&0&\widetilde{q_{31}}&0&*&\widetilde{q_{12}}\\
0&0&0&q&0&\widetilde{q_{42}}&\widetilde{q_{43}}&*
\end{bmatrix}
\end{array},
\end{align}
where $0$ correspond to events that may not occur, the rates of transitions from being dependent to being independent are constant and set to $q$, the ondiagonals, marked with $*$, are given by ${A}_{i,i}=-\sum_{j\not= i}{A}_{i,j}$, for all $i,j\in\mathcal{S}$.

We note that in general, genes are gained or lost at some Poisson rates $\alpha_k$ and $\beta_k$, per gene $k$. Therefore, we assume that, when in one of the `indepdendent-genes' states in the set $\mathcal{S}_{ind}$, we have $\alpha_1=q_{13}=q_{24}$ and $\alpha_2=q_{12}=q_{34}$, which are some Poisson rates at which we may gain gene $1$ and gene $2$, respectively; while $\beta_1=q_{31}=q_{42}$ and $\beta_2=q_{21}=q_{43}$, which are some Poisson rates at which we may lose gene $1$ and gene $2$, respectively.

Furthermore, we assume that the transition from independence to dependence may only occur from state $11$ to state $\widetilde{11}$, at some rate denoted $q_4=\lambda$; and the transition from dependence to independence occurs at some rate denoted $q=\gamma$.

Finally, we denote by $r_{00},r_{01},r_{10},r_{11},\bar{r}_{01},\bar{r}_{10}> 0$ the parameters which affect the transition rates between states within the set $\mathcal{S}_{d}$, so that given some state $i\in\mathcal{S}_{d}$, gene $k=1,2$ is gained at rate $\alpha_k r_i$ or lost at rate $\beta_k \bar{r}_i$.

By the above assumptions, the generator matrix ${\bf A}$ of the resulting process, referred to as Model I, is given by  
\begin{align}
\label{transi2}
{\bf A} =
\begin{array}{cc}&
\begin{matrix}\hspace{-0.6 cm}
00 & 01 & 10 &11 &\quad\widetilde{00} &\quad\widetilde{01} &\quad\widetilde{10} &\quad\widetilde{11}\\
\end{matrix}\\ 
\begin{matrix}
{00} \\{01} \\{10} \\{11}\\
\widetilde{00} \\\widetilde{01} \\\widetilde{10} \\ \widetilde{11}
\end{matrix}& \hspace{-0.3 cm}
\begin{bmatrix}[cccc|cccc]
*&\alpha_2&\alpha_1&0&0&0&0&0\\
\beta_2&*&0&\alpha_1&0&0&0&0\\
\beta_1&0&*&\alpha_2&0&0&0&0\\
0&\beta_1&\beta_2&*&0&0&0&\lambda\\
\hline
\gamma&0&0&0&*&\alpha_2r_{00}&\alpha_1r_{00}&0\\
0&\gamma&0&0&\beta_2r_{01}&*&0&\alpha_1\bar{r}_{01}\\	
0&0&\gamma&0&\beta_1r_{10}&0&*&\alpha_2\bar{r}_{10}\\
0&0&0&\gamma&0&\beta_1r_{11}&\beta_2r_{11}&*
\end{bmatrix}.
\end{array}
\end{align}

When two genes become associated i.e. both are needed to perform some useful function, then a transition from state $11$ to state $\widetilde{11}$ occurs. Since the function is useful, we assume that the rate of losing a gene, when both genes are present ($\widetilde{11}$ moves to $\widetilde{10}$ or  $\widetilde{01}$), would be less than the rate of losing a gene when one of the genes is already absent ($\widetilde{10}$ moves to $\widetilde{00}$, or $\widetilde{01}$ moves to $\widetilde{00}$). Therefore, $r_{11}< r_{01}, r_{10}, \bar{r}_{01}, \bar{r}_{10}$.

In Figure~\ref{fig:simtree} below, we present an example of a phylogenetic tree. To generate this tree, we simulated a tree with $50$ tips, under a pure birth-death process, where the birth rate at which speciation occurs is $1$, and the death rate at which a species becomes extinct, is $0.7$. We then randomly generated presence or absence of genes, for each species on the tree, using matrix ${\bf A}$ of Model I, given in~\eqref{transi2}. We will fit the parameters of our model to data, such as in this example, and perform the analysis, in our future work.

\begin{figure}
	\centering
	\includegraphics[width=8.5 cm, height=9 cm]{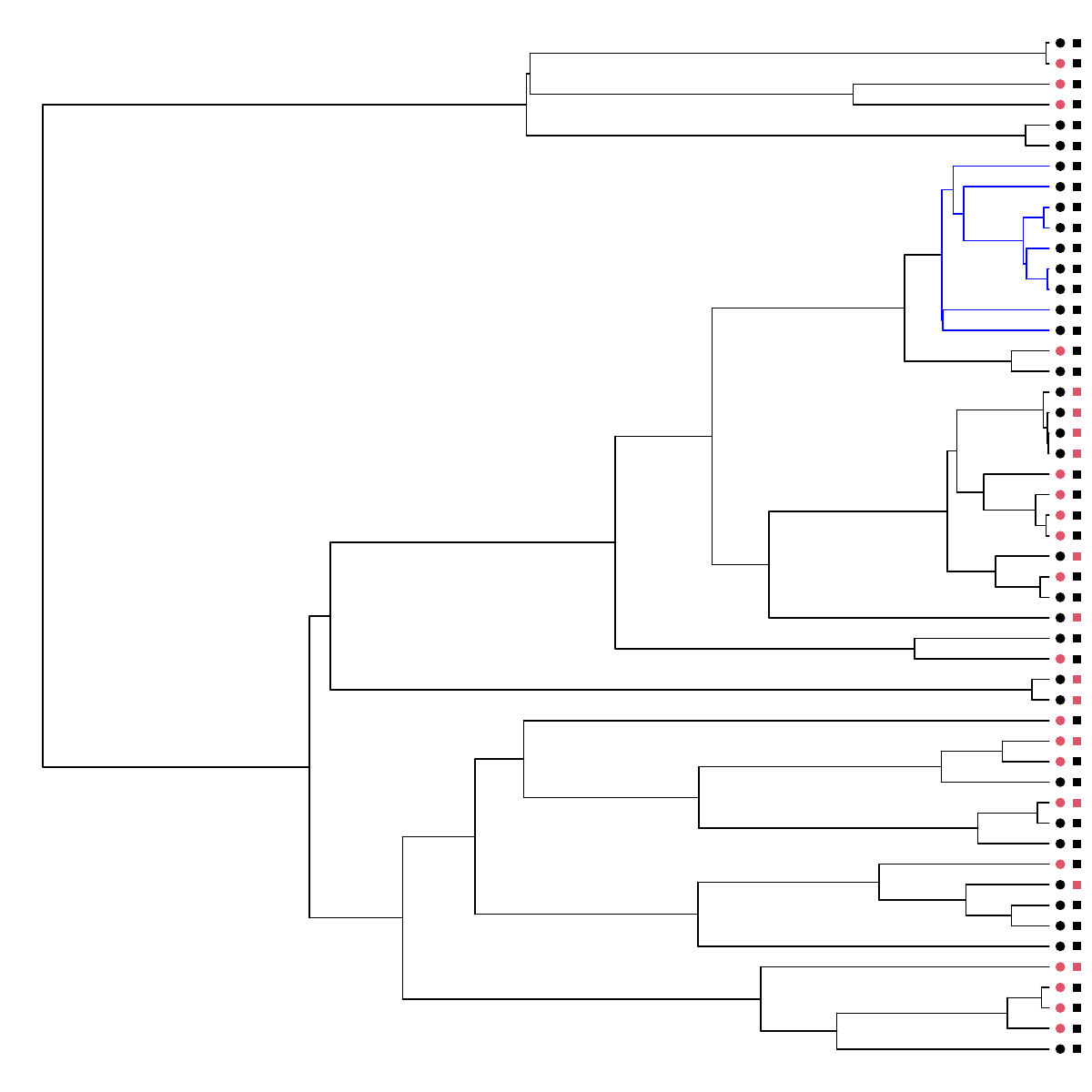}
	\caption{As an example, suppose that there are two genes in the family. Each may be in one of the two possible states: gene $1$ (circle) and gene $2$ (square), is either absent (red) or present (black) in the species (at the tip of the tree), respectively. Black lines on the tree correspond to the parts of the tree where the genes are independent of each other, while blue lines indicate dependence of the two genes.}
	\label{fig:simtree}
\end{figure}

\section{Neofunctionalisation model II}

We extend the idea of Model I in Section~\ref{sec:model1} to the following more general model of neofunctionalisation, referred to as Model II. Given $n$ potential genes that may interact with one another, let
\begin{eqnarray*}
	\mathcal{S}^{(n)}&=&\mathcal{S}^{(n)}_{ind}\cup\mathcal{S}^{(n)}_d,\\
\mathcal{S}^{(n)}_{ind}&=&\{(0;i_1,\ldots,i_n):i_k\in\{0,1\},k=1,\ldots,n\},\\
\mathcal{S}^{(n)}_d&=&\{{(1;i_1,\ldots,i_n)}:i_k\in\{0,1\},k=1,\ldots,n\},
\end{eqnarray*}
where $i_k=1,0$ corresponds to the presence or absence of gene $k$, respectively; while $(1;i_1,\ldots,i_n)$ and $(0;i_1,\ldots,i_n)$ denotes dependence and independence of the genes.

Assume as before that, given state $s=(0;i_1,\ldots,i_n)$, gene $k=1,\ldots,n$ may be gained or lost at a Poisson rate $\alpha_k$ and $\beta_k$, respectively. On the other hand, given state $s=(1;i_1,\ldots,i_n)$, gene $k=1,\ldots,n$ may be gained or lost at a Poisson rate $\alpha_k \bar r_s$ and $\beta_k r_s$, respectively.

Suppose that an additional gene may be added to the functional pathway at some Poisson rate $\delta>0$. Then, given current state $(m;i_1,\ldots,i_n)$, a transition to some state $(m;i_1,\ldots,i_n,1)$ with $i_{n+1}=1$ occurs at rate $\delta$, for $m=0,1$.

Then, to model the evolution of genes we construct a level-dependent Quasi-Birth-and-Death process (LD-QBD) $\{(X(t),\varphi(t)):t\geq 0\}$ with state space
\begin{eqnarray}
\mathcal{S}=\{(n,\varphi):n=0,1,\ldots;\varphi\in \mathcal{S}^{(n)}
\},
\end{eqnarray}
in which the level $n$ is the number of genes that may potentially interact in some functional pathway, and phase $\varphi$ records the information about the presence or absence of the genes, as well as dependence or independence.

Furthermore, assuming that $N$ is some maximum number of potential genes involved in an interaction, the generator of the process $\{(X(t),\varphi(t)):t\geq 0\}$ is matrix 
\begin{eqnarray}\label{eq:GenQ}
	\lefteqn{
	{\bf Q} 
		=
		[{\bf Q}^{[n,n^{'}]}]_{n,n^{'}\in\{2,3,\ldots,N\}}
	}
\nonumber	\\
	&=&
	\begin{bmatrix}
		{\bf Q}^{[2,2]} & {\bf Q}^{[2,3]} & {\bf 0} & \cdots &  \cdots & {\bf 0}\\
		{\bf 0} & {\bf Q}^{[3,3]} & {\bf Q}^{[3,4]} & \cdots & \cdots & {\bf 0}\\
		\vdots & \vdots & \vdots & \cdots  & \cdots & \vdots\\
		{\bf 0} & {\bf 0} & {\bf 0} & \cdots & {\bf 0} & {\bf Q}^{[N,N]}
	\end{bmatrix},
	\nonumber\\
\end{eqnarray}
with the following expressions for the block matrices in ${\bf Q}$.

For $n=2,\ldots,N-1$, the block matrix
\begin{eqnarray*}
{\bf Q}^{[n,n+1]}&=&[q_{(m;i_1,\ldots,i_n)(m';i'_1,\ldots,i'_{n+1})}]
\end{eqnarray*}  
is such that when $(m;i_1,\ldots,i_n)=(1;1,\ldots,1)$ then
\begin{eqnarray*}
q_{(m;i_1,\ldots,i_n)(m;i_1,\ldots,i_n,1)}&=&\delta
\end{eqnarray*}
and 
\begin{eqnarray*}
q_{(m;i_1,\ldots,i_n)(m';i'_1,\ldots,i'_{n+1})}&=&
0
\end{eqnarray*}
otherwise.

For $2,\ldots,N$, the block matrix
\begin{eqnarray*}
	{\bf Q}^{[n,n]}&=&[q_{(m;i_1,\ldots,i_n)(m';i'_1,\ldots,i'_n)}]
\end{eqnarray*} 
is such that 
\begin{eqnarray}
	\lefteqn{
	{\bf Q}^{[n,n]}
		=
		[{\bf Q}^{[n,n]}_{[\ell,\ell']}]_{\ell,\ell^{'}\in\{0,1,\ldots,n\}}
	}
\nonumber	\\
	&=&
	\begin{bmatrix}
		{\bf Q}^{[n,n]}_{[0,0]} & {\bf Q}^{[n,n]}_{[0,1]} & {\bf 0} & \cdots &  \cdots & {\bf 0}\\[1ex]
		{\bf Q}^{[n,n]}_{[1,0]} & {\bf Q}^{[n,n]}_{[1,1]} & {\bf Q}^{[n,n]}_{[1,2]} & \cdots & \cdots & {\bf 0}\\[1ex]
		\vdots & \vdots & \vdots & \cdots  & \cdots & \vdots\\[1ex]
		{\bf 0} & {\bf 0} & {\bf 0} & \cdots & {\bf Q}^{[n,n]}_{[n-1,n]} & {\bf Q}^{[n,n]}_{[n,n]}
	\end{bmatrix},
	\nonumber\\
\end{eqnarray}
where the block matrix
\begin{eqnarray*}
{\bf Q}^{[n,n]}_{[\ell,\ell']}&=&[q_{(m;i_1,\ldots,i_n)(m';i'_1,\ldots,i'_n)}]_{i_1+\ldots+i_n=\ell;i'_1+\ldots+i'_n=\ell'}
\end{eqnarray*}
collects all transition rates from states with $\ell$ present genes ($i_1+\ldots+i_n=\ell$) to states with with $\ell'$ present genes ($i'_1+\ldots+i'_n=\ell'$). 

As a simple example, matrix ${\bf A}$ in~\eqref{transi2} corresponds to $n=2$, and so, when rearranged, is equivalent to matrix ${\bf Q}^{[2,2]}$, given by
\begin{eqnarray*}
{\bf Q}^{[2,2]}&=&
[{\bf Q}^{[2,2]}_{[\ell,\ell']}]_{\ell,\ell^{'}\in\{0,1,2\}}
\end{eqnarray*}
with
\begin{eqnarray*}
{\bf Q}^{[2,2]}_{[0,0]}&=&
\left[
\begin{array}{cc}
*&0\\
\gamma&*
\end{array}
\right],
\\ 
{\bf Q}^{[2,2]}_{[0,1]}&=&
\left[
\begin{array}{cccc}
\alpha_2&\alpha_1&0&0\\
0&0&\alpha_2r_{00}&\alpha_1r_{00}
\end{array}
\right],
\end{eqnarray*}
and
\begin{eqnarray*}
	{\bf Q}^{[2,2]}_{[1,0]}&=&
	\left[
	\begin{array}{cccc}
\beta_2&0\\
\beta_1&0\\
0&\beta_2r_{01}\\
0&\beta_1r_{10}
	\end{array}
	\right],
\\
	{\bf Q}^{[2,2]}_{[1,1]}&=&
	\left[
	\begin{array}{cccc}
		*&0&0&0\\
		0&*&0&0\\
		\gamma&0&*&0\\
		0&\gamma&0&*
	\end{array}
	\right],\\
	{\bf Q}^{[2,2]}_{[1,2]}&=&
	\left[
	\begin{array}{cc}
\alpha_1&0\\
\alpha_2&0\\
0&\alpha_1\bar r_{01}\\
0&\alpha_2 \bar r_{10}
	\end{array}
	\right],
\end{eqnarray*}
and
\begin{eqnarray*}
	{\bf Q}^{[2,2]}_{[2,1]}
	&=&
	\left[
	\begin{array}{cccc}
\beta_1&\beta_2&0&0\\
0&0&\beta_1r_{11}&\beta_2r_{11}
	\end{array}
	\right],
	\\ 
	{\bf Q}^{[2,2]}_{[2,2]}&=&
	\left[
	\begin{array}{cc}
*&\lambda\\
\gamma&*
	\end{array}
	\right].
\end{eqnarray*}

That is,
\begin{eqnarray*}
	\lefteqn{
{\bf Q}^{[2,2]}
}
\\
&=&\left[
\begin{array}{ccc}
	{\bf Q}^{[2,2]}_{[0,0]}&{\bf Q}^{[2,2]}_{[0,1]}&{\bf 0}\\[2ex]
	{\bf Q}^{[2,2]}_{[1,0]}&{\bf Q}^{[2,2]}_{[1,1]}&{\bf Q}^{[2,2]}_{[1,2]}\\[2ex]
	{\bf 0}&{\bf Q}^{[2,2]}_{[2,1]}&{\bf Q}^{[2,2]}_{[2,2]}\\[2ex]
\end{array}
\right]
\\
&=&
\left[
\begin{array}{cc|cccc|cc}
*&0&\alpha_2&\alpha_1&0&0&0&0\\
\gamma&*&0&0&\alpha_2r_{00}&\alpha_1r_{00}&0&0\\\hline
\beta_2&0&*&0&0&0&\alpha_1&0\\
\beta_1&0&0&*&0&0&\alpha_2&0\\
0&\beta_2r_{01}&\gamma&0&*&0&0&\alpha_1\bar r_{01}\\
0&\beta_1r_{10}&0&\gamma&0&*&0&\alpha_2\bar r_{10}\\\hline
0&0&\beta_1&\beta_2&0&0&*&\lambda\\
0&0&0&0&\beta_0r_{11}&\beta_2r_{11}&\gamma&*
\end{array}
\right].
\end{eqnarray*}

If the maximum number of potential genes involved in an interaction is $N=2$, then we simply have 
\begin{eqnarray}
{\bf Q}&=&{\bf Q}^{[2,2]},
\end{eqnarray}
however, if $N>2$, then ${\bf Q}$ has the more general form, as in~\eqref{eq:GenQ}, and then ${\bf Q}^{[2,3]}$ is an $8\times 16$ matrix, with
\begin{eqnarray}
	{\bf Q}^{[2,3]}
	&=&
	\left[
	\begin{array}{cc}
		{\bf 0}&{\bf 0}\\
		{\bf 0}&\delta
	\end{array}
	\right],
\end{eqnarray}
since the only possible transition in such matrix is from state $(m;i_1,\ldots,i_n)=(1;1,\ldots,1)$  to state $(m';i'_1,\ldots,i'_{n+1})=(1;1,\ldots,1,1)$, which occurs at rate $\delta$.

\bigskip
The expressions for various metrics of interest of such defined LD-QBD can then be obtained by applying results from the literature of matrix-analytic methods, including e.g. Ramaswami~\cite{ram1997}, Joyner and Fralix~\cite{joyner2016new}, Phung-Duc et al.~\cite{phung2010simple} and Samuelson et al.~\cite{SOB2020}. Therefore, the analysis of the evolution of gene content and their association across a phylogeny is possible with our model proposed here. 

\bibliographystyle{abbrv}
\bibliography{sigproc_arxiv}
\end{document}